\documentstyle[12pt]{article}

\input{epsf.tex}

\begin{document}

\title{Remarks on a solvable cosmological model}

\author{Ion I. Cot\u aescu \thanks{E-mail:~~~cota@physics.uvt.ro}\\ 
and\\
Dumitru N. Vulcanov \thanks{E-mail:~~~vulcan@physics.uvt.ro}\\
{\small \it West University of Timi\c soara,}\\
       {\small \it V. P\^ arvan Ave. 4, RO-1900 Timi\c soara, Romania}}

\date{\today}

\maketitle

\begin{abstract}

We present an exact analytical solution of the Einstein
equations with cosmological constant in a spatially flat Robertson-Walker 
metric. This is interpreted as an isotropic Lemaitre-type version of the 
cosmological Friedmann model. Implications in the recent discovered cosmic 
acceleration of the universe and in the theory of an inflationary model of 
the universe are in view. Some properties of this solution are  pointed out 
as a result of numerical investigations of the model.

Pacs 04.62.+v

\end{abstract}

\newpage

\section{Introduction}

Recent astrophysical investigations \cite{2,3} demonstrate that the 
expansion of the universe is accelerating rather than slowing down. This may 
change our picture of the universe suggesting that we live in an  
accelerating and flat universe. However, from the theoretical point of view 
intensive efforts are done \cite{4} in order to accommodate the relativistic 
cosmology with an accelerating universe. One of the major results of this 
situation is the fact that the most popular solution (inspired by what is 
happening in the theory of inflationary cosmology) is to consider models which 
satisfy Einstein equations  with cosmological constant in order to
induce cosmic acceleration. Actually, several models until now not being in the
main stream of the modern cosmology, are again in view of cosmologists.
We believe that it is necessary to more carefully 
investigate new possible versions of the the Friedman-Robertson-Walker (FRW) 
model \cite{5,6} in order to see how may we implement here cosmic acceleration
without loosing the well-known results of the model in describing early stages 
of the universe. 

Another important problem is to try to find simple analytically solvable models
which may be easily interpreted from the physical point of view.
Moreover, it is clear that many further developments of the more complicated 
models have to be done using algebraic or numerical methods on computers. Then 
it is important to have at least one simple analytically solvable model for  
testing the computational methods and verify their degree of confidence. 

This article is dedicated to the revealing of an analytical solution of the 
Einstein equations with cosmological constant in a version of the FRW model 
with flat space metric \cite{5}. We present
some  properties of this solutions and the possible implications to the
more accurate description of the inflationary and accelerated universe. 

We must specify that the model with cosmological constant and $k=1$ is known 
in literature as  Lemaitre-FRW model but, unfortunately, this does not have 
analytical solution. In other respects, we believe that the solution for $k=0$ 
we present here is ignored by the investigators of this cosmology since  
they are forced to find solutions of the Einstein equations in terms of the 
measurable 
quantities (as the Hubble constant/parameter, the redshift or the deceleration 
parameter) and searching for an approximate solution for the scale factor of 
the universe as a Taylor expansion in terms of these quantities. Thus the 
existence of an analytical solution is hidden by the large number of solutions 
available only for  restricted periods from the history of universe.

\section{The model and its solution}

Let us consider the Lemaitre-type version of the FRW model with 
flat space (i.e., $k=0$) \cite{5}, positively defined cosmological constant, 
$\Lambda$, 
and linear dependence between the density of energy, $\epsilon$, and the 
pressure, $p$. In general, in an arbitrary local chart (or natural frame) of 
coordinates $x^{\mu}$ ($\mu,\nu,..=0,1,2,3$), the Einstein equations of this 
model (written in natural units with $c=1$), 
\begin{equation}
R_{\mu}^{\nu}-\frac{1}{2}R\delta^{\nu}_{\mu}-\Lambda \delta _{\mu}^{\nu}
=8\pi G T^{\nu}_{\mu}\,,
\end{equation}
involve the usual classical stress-energy tensor 
\begin{equation}
T_{\mu}^{\nu}=(\epsilon+p)u^{\nu}u_{\mu}-p \delta_{\mu}^{\nu}\,, 
\end{equation}
which depends on the covariant four-velocity  $u^{\mu}=dx^{\mu}/ds$.

In the preferred frame with Cartesian coordinates $x^0=t,\, \vec{x}$ 
which has the usual FRW line element with  $k=0$,  
\begin{equation}\label{le}
ds^2=dt^2 - a(t)^2d\vec{x}^2\,,    
\end{equation} 
we have $u^0=u_0=1$ and $u^i=0$ ($i=1,2,3$) such that the Einstein equations 
reduce to the simple system
\begin{eqnarray}
3\left(\frac{\dot a}{a}\right)^2&=&\Lambda + 8\pi G \,\epsilon \label{E1}\\
2\frac{\ddot a}{a}
+\left(\frac{\dot a}{a}\right)^2 &=&\Lambda - 8\pi G\, p \label{E2}
\end{eqnarray}
where we denoted $\dot{~}=d/dt$. Assuming that $\epsilon$ and $p$ are 
functions only on $t$ and satisfy a linear equation of state, namely
\begin{equation}\label{state}
p=\kappa \epsilon\,,
\end{equation} 
one can integrate this system with the conditions 
\begin{equation}\label{ini} 
\epsilon(0)=\epsilon_0\,, \quad a(0)=1\,,
\end{equation}
at the present time $t=0$.    
Indeed, (\ref{E1}), (\ref{E2}) and (\ref{state}) are equivalent with 
the equation 
\begin{equation}
 \frac{d}{dt}\left(\frac{\dot a}{a}\right)=\frac{\kappa+1}{2}
\left[\Lambda-3\left(\frac{\dot a}{a}\right)^2\right]
\end{equation}   
which, after a little calculation and exploiting (\ref{ini}), leads to 
the final solution
\begin{eqnarray}
a(t)&=&\left[{\rm cosh}\frac{\sqrt{3\Lambda}}{2}(\kappa+1)\,t 
\right.\nonumber\\  
&&~~\left.+\sqrt{1+\frac{8\pi G\,\epsilon_{0}}{\Lambda}}\,
{\rm sinh}\frac{\sqrt{3\Lambda}}{2}(\kappa+1)\,t\right]^{2/3(\kappa+1)}   
\end{eqnarray}
while  the density of energy reads
\begin{equation}\label{dens}
\epsilon(t)=\epsilon_0\,a(t)^{-3(\kappa+1)}\,.
\end{equation}
Finally, we calculate the Bang time, $-t_0$, from the initial condition 
$a(-t_{0})=0$ as 
\begin{equation}\label{BB-time}
t_0=\frac{2}{\sqrt{3\Lambda}\,(\kappa+1)}{\rm argcoth}\sqrt{1+
\frac{8\pi G\,\epsilon_{0}}{\Lambda}}\,,
\end{equation} 
completing  thus the solution of this simple model. 

Let us observe that the conservation law $T^{\mu}_{\nu ; \mu} = 0$ is 
implicitly fulfilled since  (\ref{dens}) is simultaneously the
solution of this conservation equation which takes the form 
\begin{equation}\label{conserv}
\dot{\epsilon} + 3 \epsilon (\kappa+1)\frac{\dot{a}}{a}=0
\end{equation}
when the equation of state (\ref{state}) is accomplished. Moreover, 
it is important that the equation (\ref{dens}) is independent 
on $\Lambda$ since then it holds even in the particular case of $\Lambda\to 0$ 
when we recover the usual solution of the FRW model with $k=0$ \cite{5,6}, 
\begin{equation}
a(t) =
\left[ 1 + \sqrt{6\pi G\epsilon_0}(\kappa +1)\,t \right]^{2/3(\kappa+1)}.
\end{equation}

The interesting new features of our model are due to the cosmological constant 
which gives an hyperbolic character to the general model or to its asymptotic 
behavior. It is clear that the universe devoid of matter 
becomes a de Sitter one but it is remarkable that in the far future the time 
behavior of $a(t)$ is also of the de Sitter type since for very 
large $t$ we have 
\begin{equation}
a(t)\sim 
\alpha \,e^{\sqrt{\frac{\Lambda}{3}}\,t}\,, \quad \alpha= 
\left(\frac{1}{2}+\frac{1}{2}\sqrt{1+\frac{8\pi G\epsilon_{0}}{\Lambda}}
\right)^{2/3(\kappa+1)}\,.
\end{equation}
In other words if we rescale the  Cartesian space coordinates with 
the factor $\alpha$ then the line element (\ref{le}) becomes just the de Sitter
one in the limit of $t\to \infty$. 
Thus we can conclude that our solution is compatible with the well-known
models describing the Big-Bang scenario and for late behavior it
shows off characteristics of a de Sitter spacetime.

Many physical effects depend on the form of the Hubble function that in our 
model reads 
\begin{equation}\label{hubble}
H(t) = \frac{\dot{a}}{a} = \frac{1}{\sqrt{3(\Lambda+8\pi G\epsilon_0)}}
\left[ \Lambda+ \frac{8\pi G\epsilon_0}
{1 +
\sqrt{1+\frac{8\pi G\epsilon_0}{\Lambda}}{\rm tanh} \frac{\sqrt{3\Lambda}}{2}
(\kappa + 1)\,t}\right]\,.
\end{equation}
Obviously, in the late time limit we have
\begin{eqnarray}
\lim_{t \rightarrow \infty} H(t) = \sqrt{\frac{\Lambda}{3}} \nonumber
\end{eqnarray}
while the actual value of the
Hubble function  (i.e., the Hubble constant),  
\begin{equation}\label{hubble_const}
H_0 = H(t)\mid_{t=0} = \sqrt{\frac{\Lambda + 8 \pi G \epsilon_0}{3}}\,,
\end{equation}
depends on the actual value of the density of energy, $\epsilon_0$, and
$\Lambda$.

\section{Numerical investigations}

\begin{figure}
\epsfxsize=2.4in
\epsfysize=2.4in
\epsfbox{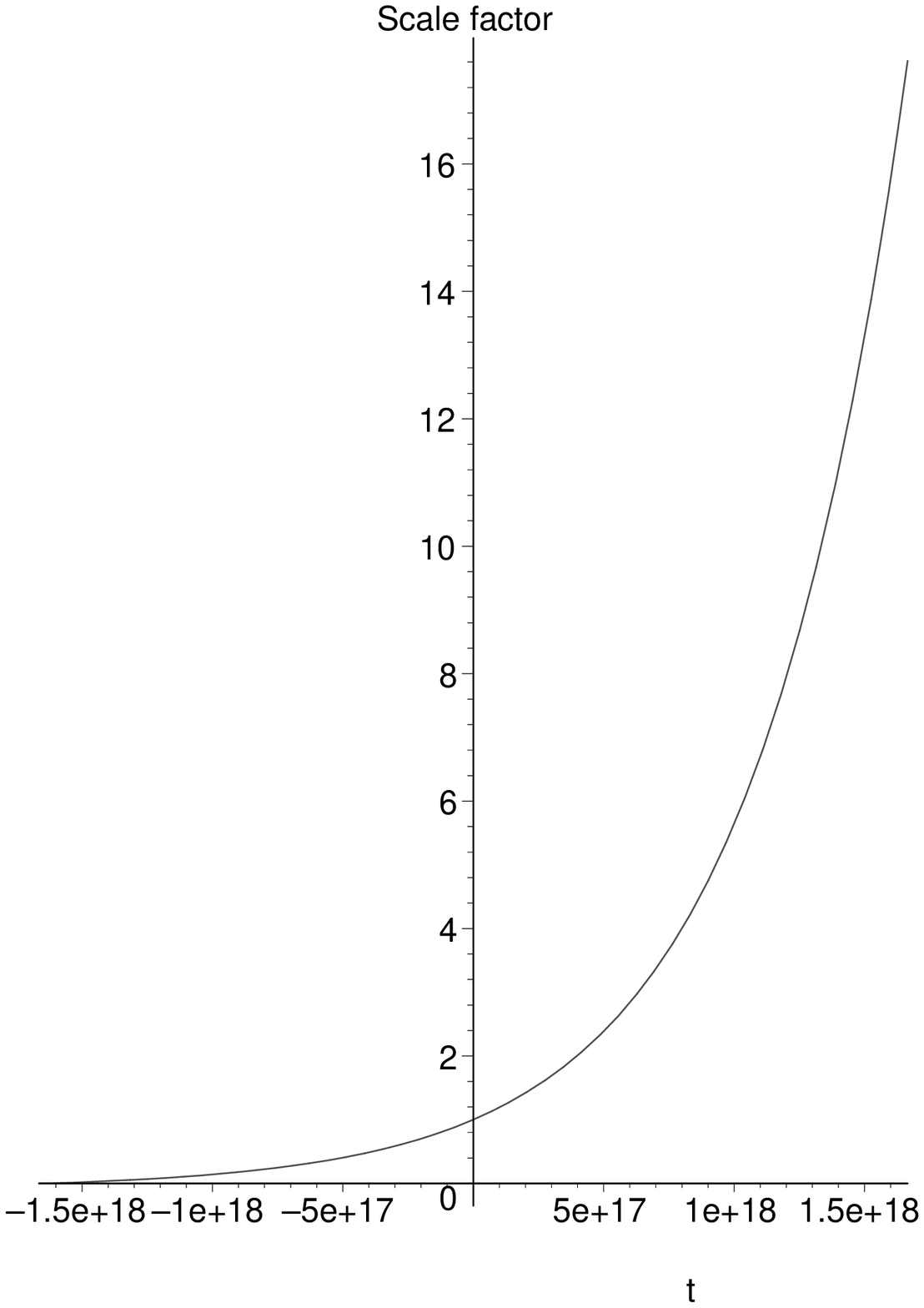}
\vspace{-2.4in}
\hspace{2.8in}
\epsfxsize=2.4in
\epsfysize=2.4in
\epsfbox{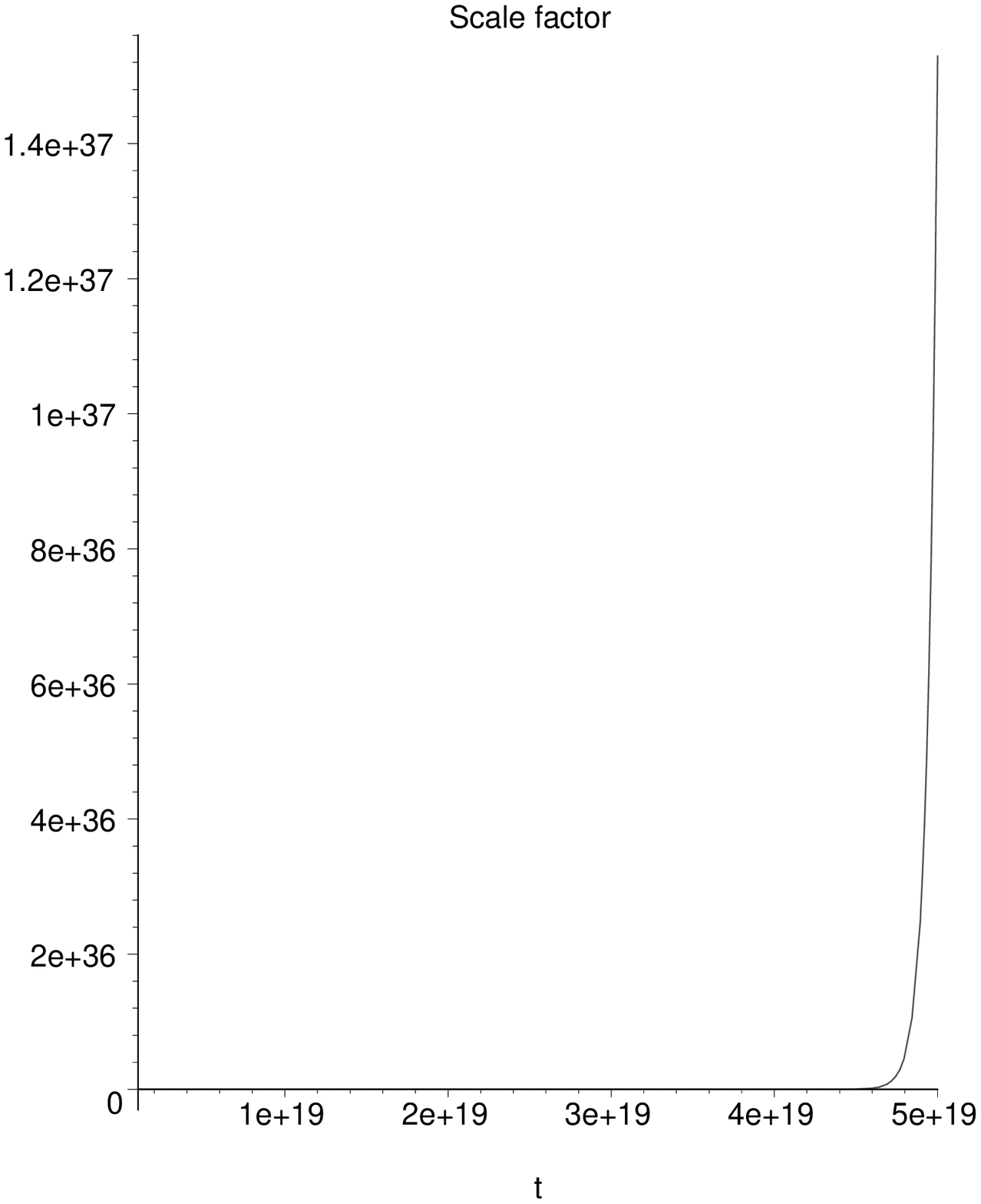}
\caption{Evolution of the scale factor $a(t)$ from the Big-Bang (left panel) 
and the fast growing of the scale factor at late time $t=10^{18}...
5\, 10^{19}~s$ (right panel)}
\label{fig:raza1}
\end{figure}

This section is dedicated to several numerical results we obtained using
our analytical solution which reveals some interesting features
of the model. First we have plotted the behavior of the scale factor
in time for a constant value of the cosmological constant $\Lambda$. In this 
purpose we considered an accepted value of the actual mass-density of the 
universe as $\rho_0=3 \cdot 10^{-28} Kg/m^3$ (in IS units where  
$\epsilon_0=\rho_0\, c^2$) and of the
actual Hubble constant as $H_0 = (1.8 \, 10^{10}~years)^{-1}$ \cite{8}.
Then, using (\ref{hubble_const}) we can estimate the cosmological constant
as being $\Lambda \approx 0.977 \, 10^{-52} ~m^{-2}$ which will be the value 
used in the next numerical investigations. We also choose $\kappa = -1/2$.
Two first graphs are showed in Figure \ref{fig:raza1} for different intervals 
of time, one before the actual time $t=0$ and the other one for late behavior 
of the universe. It is obvious, especially from the second picture 
(right panel) the fast growing of the scale factor for late time, showing 
the accelerated stage of the universe described by this model.
We have to mention that we estimated the Big-Bang time $t_0$ using the above
numerical values and formula (\ref{BB-time}). Thus we have used in our
figures the resulting value $t_0\approx 0.1665 \, 10^{19}$s ($\approx
52.7$ billion years). This value is larger than the recognized one and
appears here due to the small value of the mass-density we used. A greater
value for $\rho$, including dark matter and other unknown components,(for 
example one ten times larger) gives a Big-Bang time $t_0 \approx 29$ 
billion years.  .

\begin{figure}
\epsfxsize=2.4in
\epsfysize=2.4in
\epsfbox{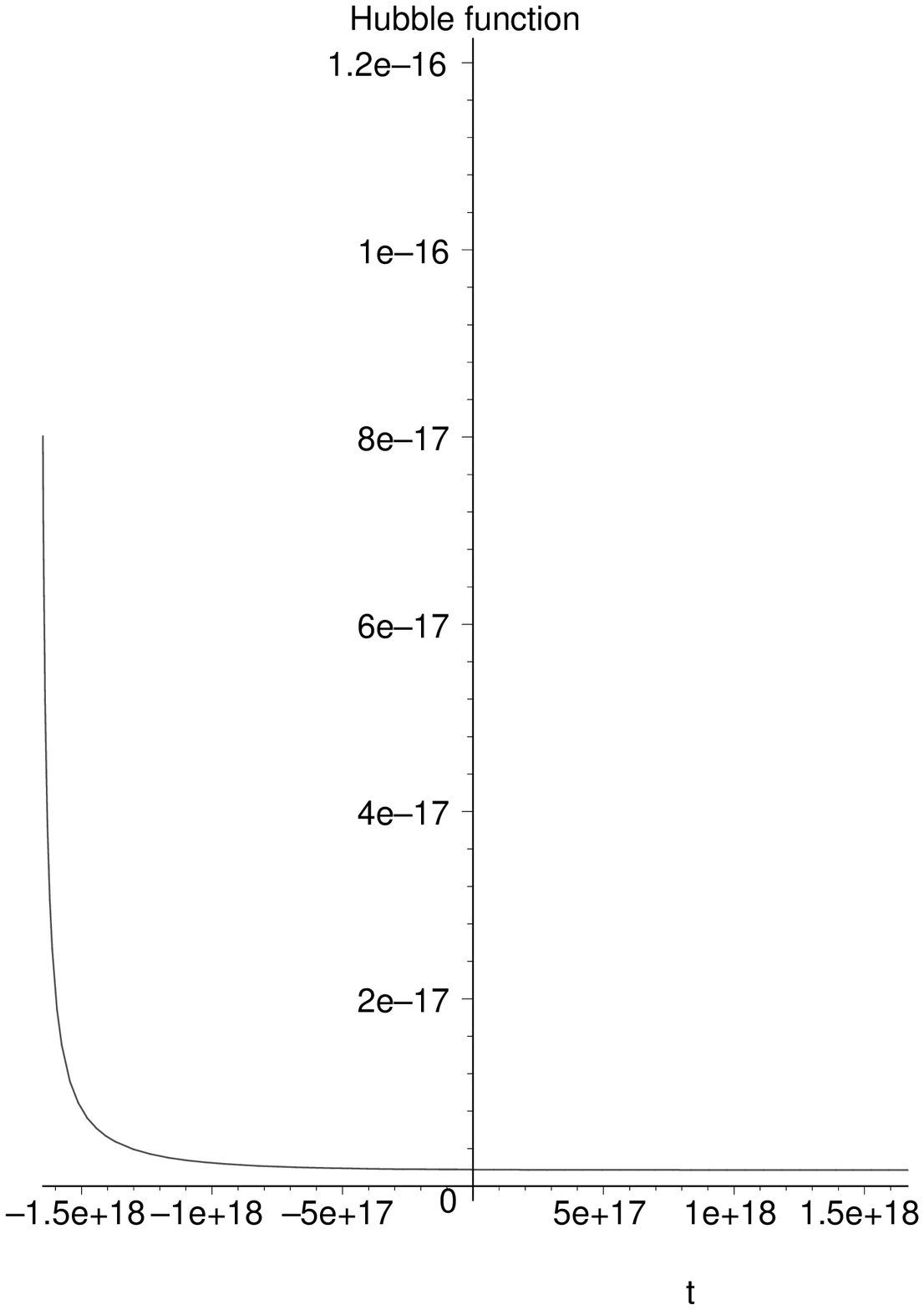}
\vspace{-2.4in}
\hspace{2.8in}
\epsfxsize=2.4in
\epsfysize=2.4in
\epsfbox{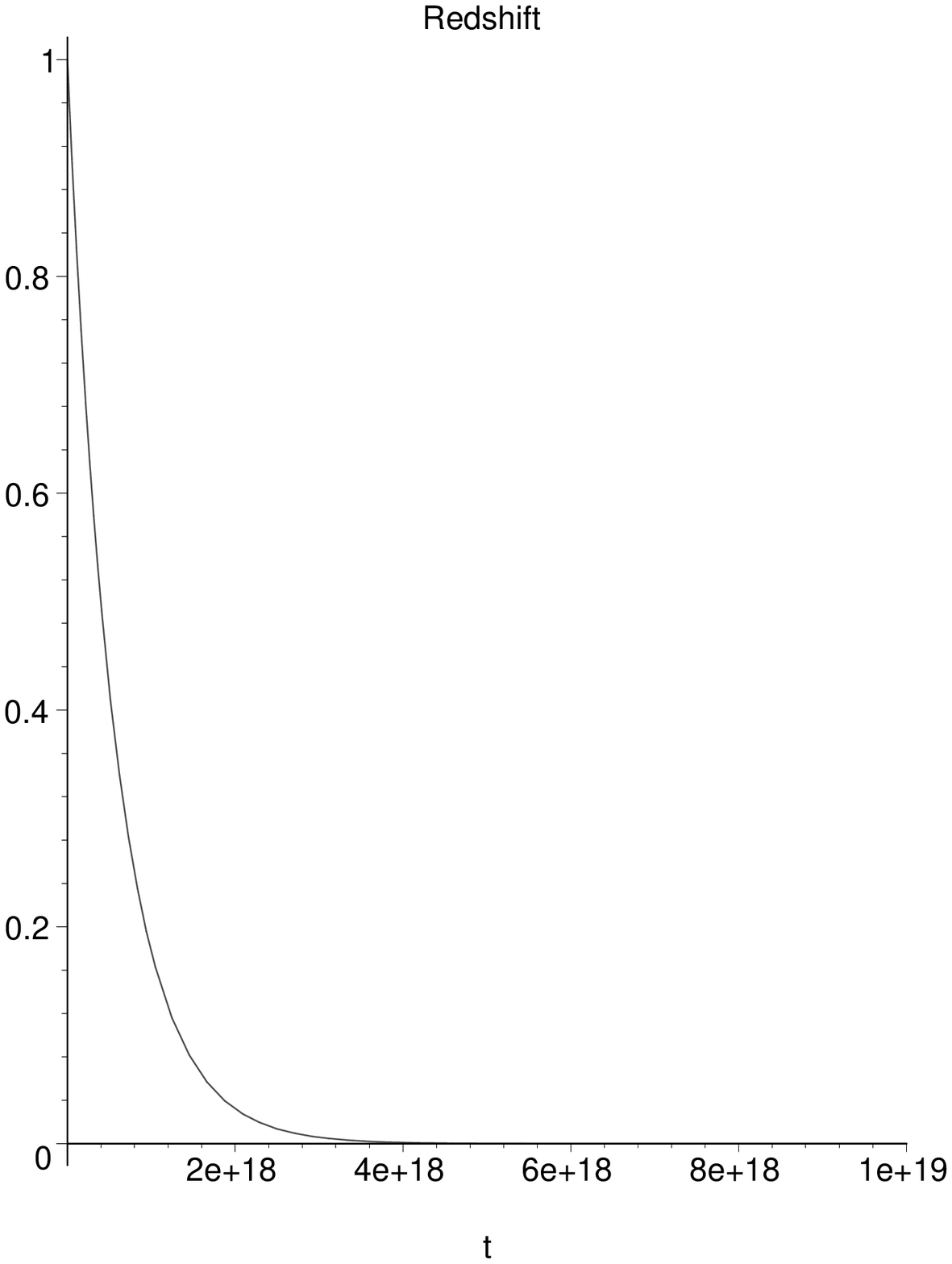}
\caption{Evolution of the Hubble function (left panel) and of the redshift 
defined as $1/a(t)$  (right panel)} 
\label{fig:hubble1}
\end{figure}

Figure  \ref{fig:hubble1} presents the time behavior of the Hubble
constant (the left panel) as defined in (\ref{hubble}) for the same 
parameters as for the above graphs. As we 
pointed out in the comments ending the last section, the Hubble function 
is evolving to a constant value $\sqrt{\Lambda / 3}$ specific for the de 
Sitter model. This conclusion is coupled to the late time behavior of 
the scale factor (see above) which is also specific to a de Sitter metric. 
In the right panel of the same Figure \ref{fig:hubble1} we plotted the
redshift defined as $1/a(t)$. 

\begin{figure}
\epsfxsize=2.4in
\epsfysize=2.4in
\epsfbox{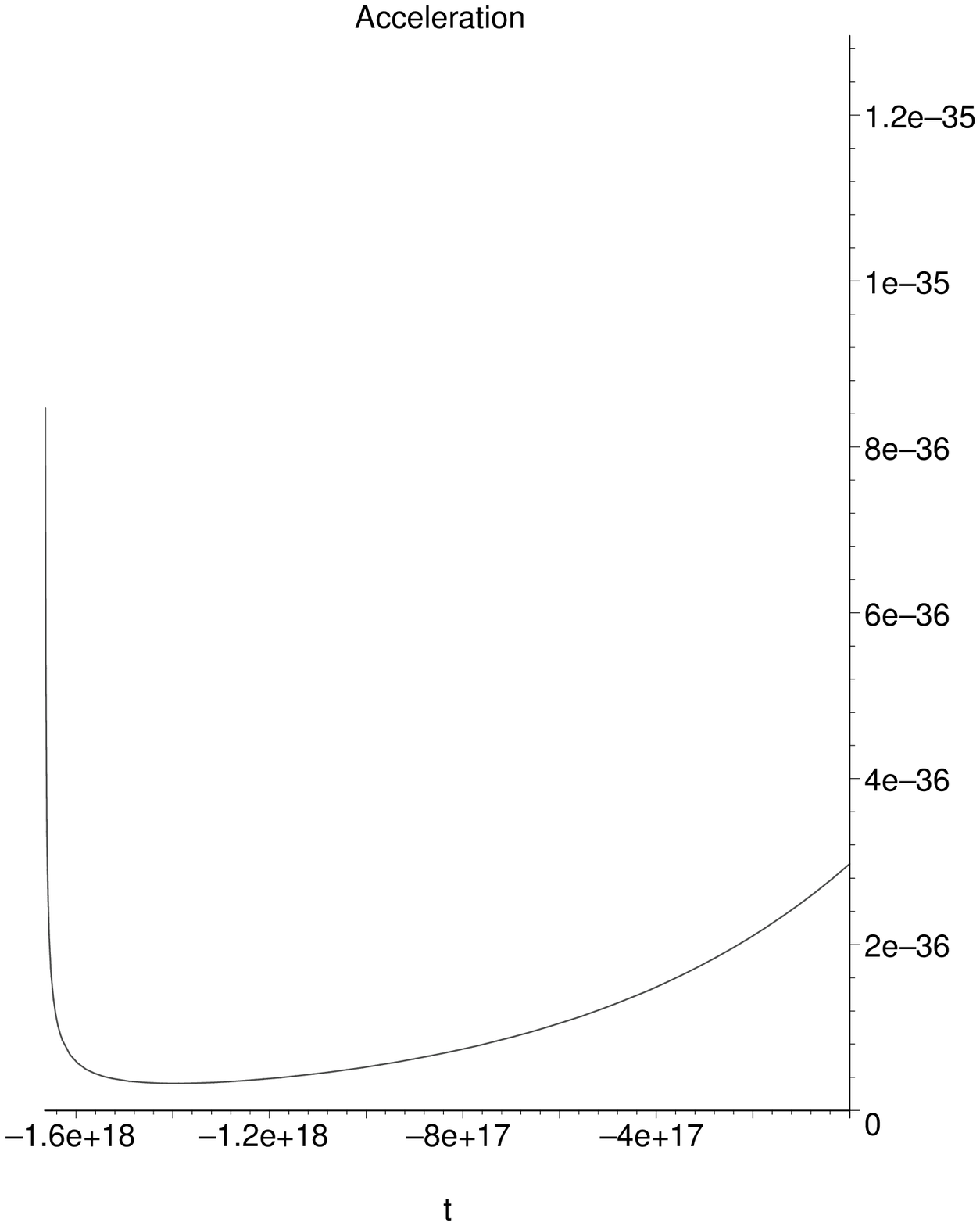}
\vspace{-2.4in}
\hspace{2.8in}
\epsfxsize=2.4in
\epsfysize=2.4in
\epsfbox{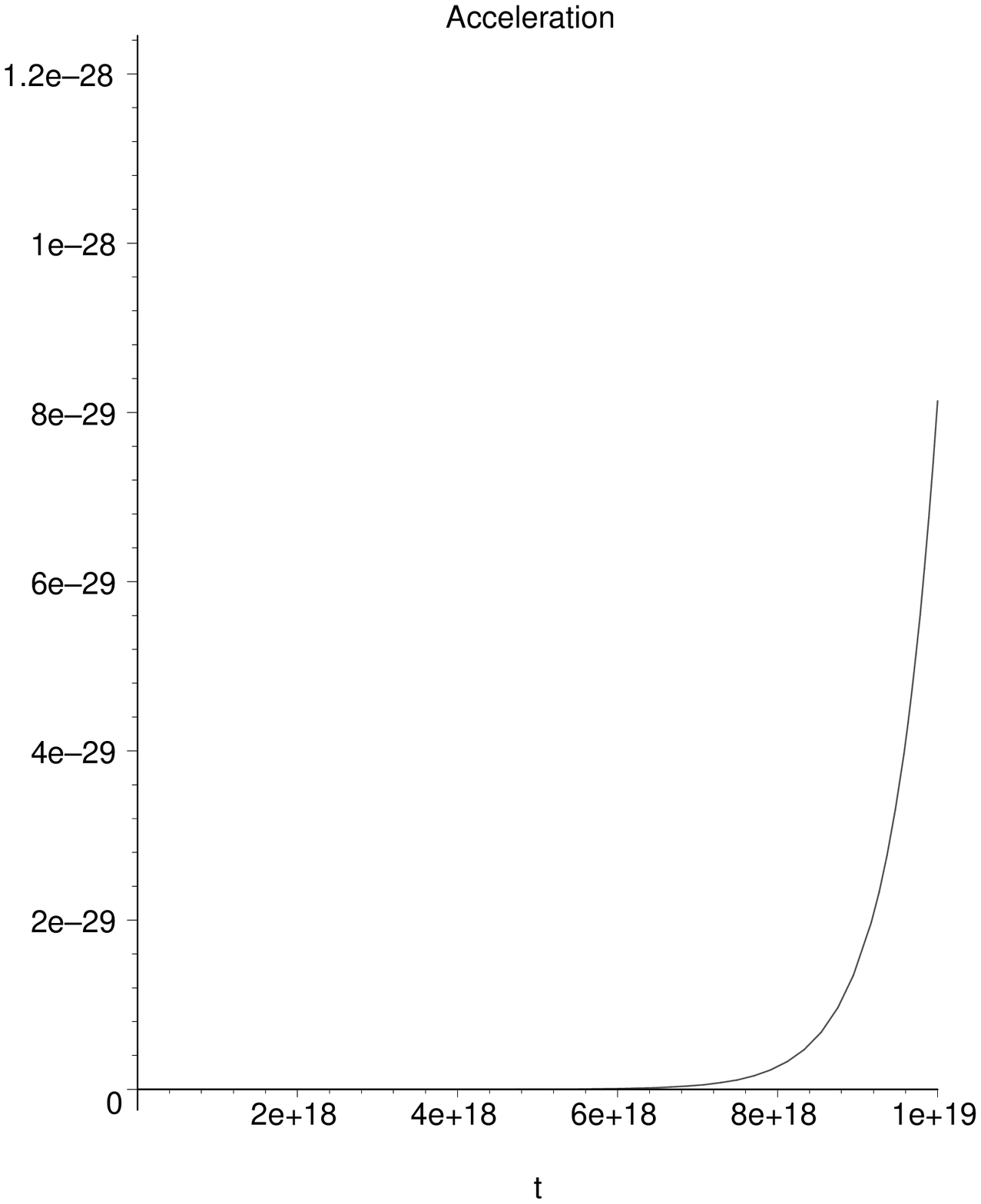}
\caption{Evolution of the acceleration, $\ddot{a}(t)$,
before the present time (left panel) and after $t=0$ (right panel)}
\label{fig:acc-acc}
\end{figure}

\begin{figure}
\epsfxsize=2.4in
\epsfysize=2.4in
\epsfbox{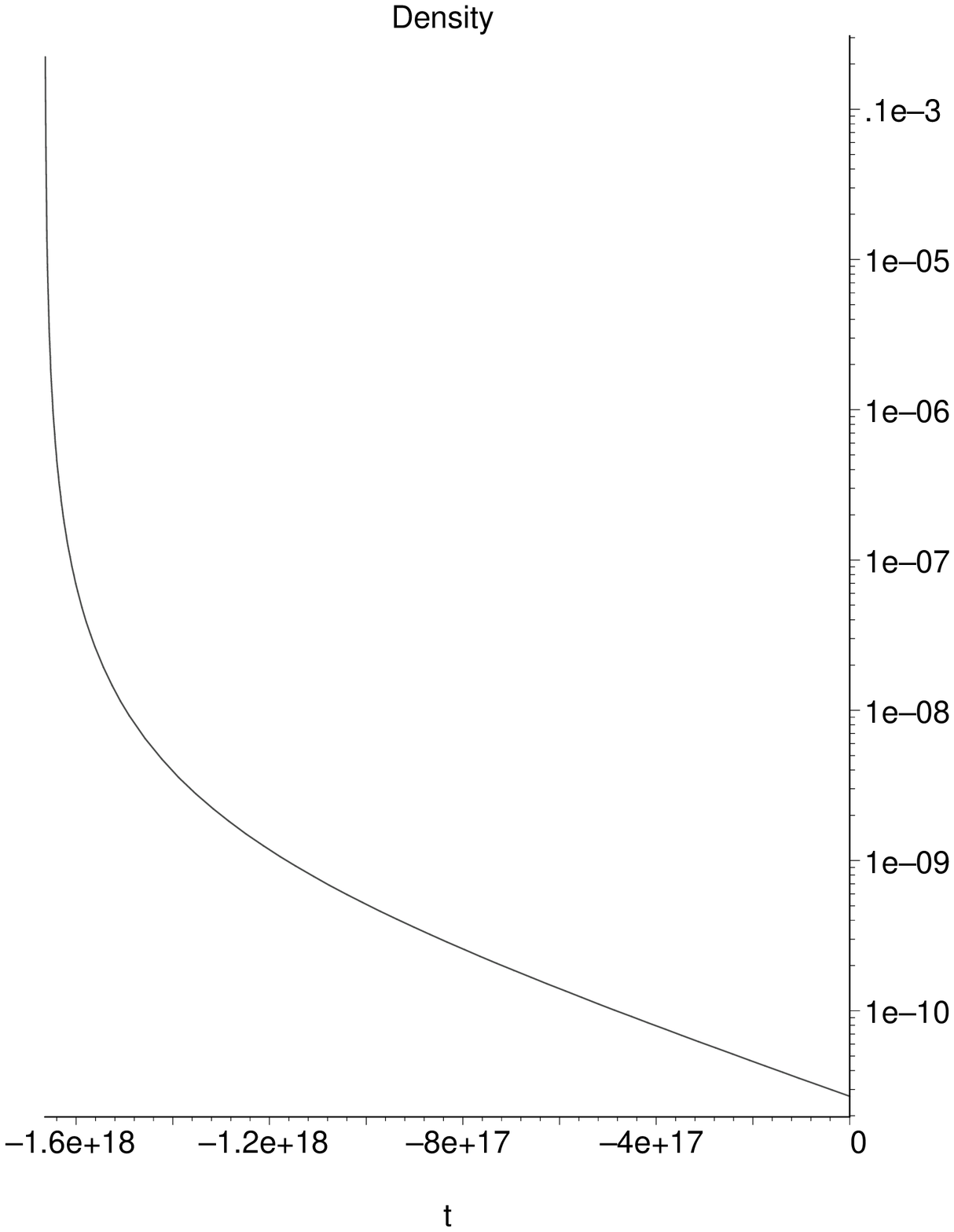}
\vspace{-2.4in}
\hspace{2.8in}
\epsfxsize=2.4in
\epsfysize=2.4in
\epsfbox{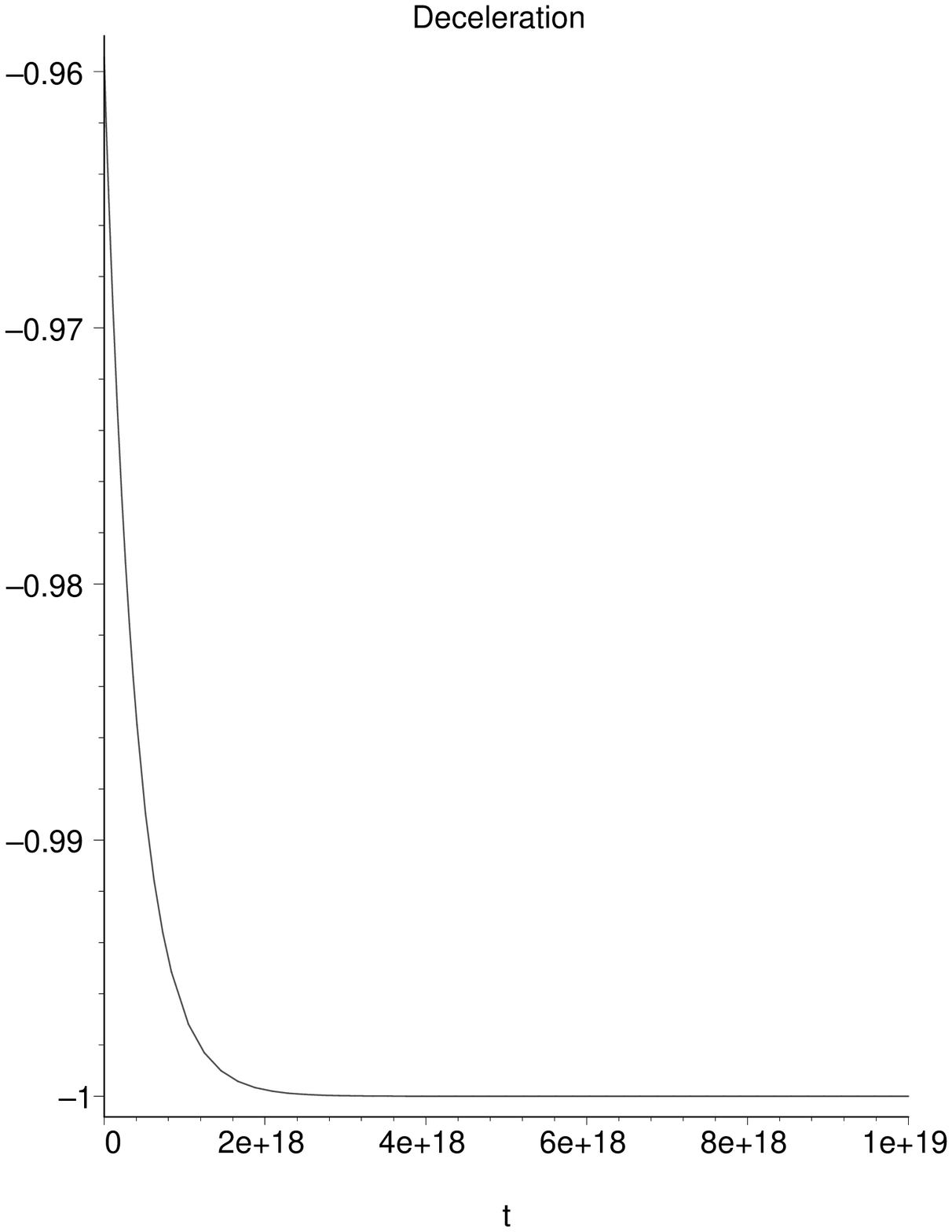}
\caption{Evolution of the density before the actual time $t=0$
(left panel) and of the deceleration, $\ddot{a}(t)/[a(t)\,
H(t)^2]$, after the actual time (right panel)}
\label{fig:dens-dec}
\end{figure}

Next plots, Figure \ref{fig:acc-acc}, are dedicated to the time 
behavior of the acceleration defined here as the second time derivative of 
the scale factor, $\ddot{a}(t)$. Before the actual time $t=0$ (presented in
the left panel) shows  the decreasing of the acceleration after the 
Big-Bang at $t \ge -t_0$. In the right panel we observe the increasing of 
the acceleration at later time in the universe evolution. 
Thus we can conclude that it will
be possible to use our solution for modeling the so-called 
``cosmic-acceleration'' in discussion in the modern astrophysics.

Last of this type of plots, Figure \ref{fig:dens-dec}, represent the time 
behavior of the density function $\epsilon(t)$  given by equation (\ref{dens})
and of the deceleration  defined as $\ddot{a}(t)/[a(t)\, H(t)^2]$   
\cite{5,8}.

\section*{Acknowledgments}

One of the authors (D. N. V.) is deeply indebted to Prof. B. F. Schutz and
Prof. E. Seidel for several invitations at the Albert Einstein Institute,
Golm, Germany where partially this research was done. Manny thanks also
for fruitful discussions on the subject with M. Alcubierre and 
F. Guzman.

\end{document}